\documentclass[sigconf,nonacm]{acmart}

\AtBeginDocument{%
  }

\begin{document}

\title{Effect of Appearance and Animation Realism on the Perception of Emotionally Expressive Virtual Humans}

\author{Nabila Amadou}
\affiliation{%
  \institution{Utrecht University}
  \city{Utrecht}
  \country{The Netherlands}}
\email{nabila.amadou@gmail.com}

\author{Kazi Injamamul Haque}
\affiliation{%
  \institution{Utrecht University}
  \city{Utrecht}
  \country{The Netherlands}}
\email{k.i.haque@uu.nl}

\author{Zerrin Yumak}
\affiliation{%
  \institution{Utrecht University}
  \city{Utrecht}
  \country{The Netherlands}}
\email{z.yumak@uu.nl}

\renewcommand{\shortauthors}{Amadou, Haque and Yumak}

\begin{abstract}
  3D Virtual Human technology is growing with several potential applications in health, education, business and telecommunications. Investigating the perception of these virtual humans can help guide to develop better and more effective applications. Recent developments show that the appearance of the virtual humans reached to a very realistic level. However, there is not yet adequate analysis on the perception of appearance and animation realism for emotionally expressive virtual humans. In this paper, we designed a user experiment and analyzed the effect of a realistic virtual human's appearance realism and animation realism in varying emotion conditions. We found that higher appearance realism and higher animation realism leads to higher social presence and higher attractiveness ratings. We also found significant effects of animation realism on perceived realism and emotion intensity levels. Our study sheds light into how appearance and animation realism effects the perception of highly realistic virtual humans in emotionally expressive scenarios and points out to future directions.
\end{abstract}

\begin{CCSXML}
<ccs2012>
   <concept>
       <concept_id>10010147.10010371.10010352</concept_id>
       <concept_desc>Computing methodologies~Animation</concept_desc>
       <concept_significance>500</concept_significance>
       </concept>
   <concept>
       <concept_id>10003120.10003121.10003124.10010865</concept_id>
       <concept_desc>Human-centered computing~Graphical user interfaces</concept_desc>
       <concept_significance>500</concept_significance>
       </concept>
 </ccs2012>
\end{CCSXML}

\ccsdesc[500]{Computing methodologies~Animation}
\ccsdesc[500]{Human-centered computing~Graphical user interfaces}

\keywords{Perception of virtual humans, Appearance and animation realism, Emotionally expressive virtual humans}
\begin{teaserfigure}
  \includegraphics[width=\textwidth]{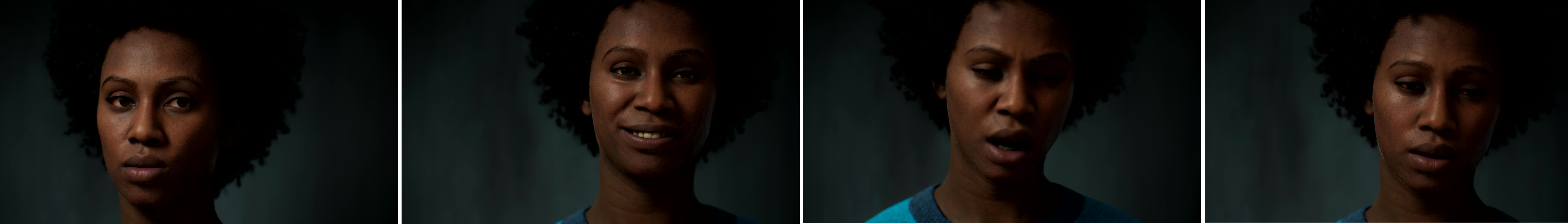}
  \caption{\textit{\textbf{Photorealistic}} appearance condition. From left to right: \textit{\textbf{Neutral, Happy, Angry, Sad}}.}
  %\Description{Screen captures of all Emotion Scenarios (here for the \textit{\textbf{Photorealistic}} appearance condition). From left to right: \textit{\textbf{Neutral, Happy, Angry, Sad}.}
  \label{fig:teaser}
\end{teaserfigure}

\maketitle

\section{Introduction}
Recent industry breakthroughs - for instance with Epic Games' MetaHumans \citep{metahumanRef}, Ziva Dynamic \citep{zivaRef} have achieved a level of appearance realism that resembles a real human. However, there is not yet enough understanding about the effects of virtual human's appearance and animation on people's perception of these characters. These analysis can help game developers and artists to make more informed choices when designing virtual humans with regards to appearance and animation quality. Recent research on perception of virtual humans has suggested that higher gesture motion realism \citep{ferstl2021}, higher appearance realism \citep{higgins_uncanny} and higher facial animation realism \citep{TINWELL20131617} leads to lower discomfort. Comparisons have been made between photorealistic and stylized human agents \citep{ferstl2021}, real humans and photorealistic agents \citep{seymour2022face} as well as photorealistic vs semi-realistic virtual humans \citep{higgins_uncanny}. However, no research has been done yet to investigate the effects of appearance realism and facial animation realism for state-of-the-art realistic emotionally expressive virtual humans.
In this paper, we investigate the effects of appearance realism and facial animation realism on the perception of a realistic virtual human and within the context of emotional scenarios. In an on-screen user study, we analyzed the effects of appearance and facial animation realism on social presence, affinity, realism and emotion intensity. The virtual human was designed with state-of-the-art realism using UnrealEngine MetaHumans \citep{metahumanRef} and Dynamixyz Facial Capture system \citep{DynamixyzRef} in varying appearance and animation realism levels. Moreover, we created different emotional scenarios. We found that higher appearance realism and higher fidelity animations led to significantly higher social presence and attractiveness ratings towards the virtual human as expected in our hypothesis. We also found significant effects of animation realism on the perceived realism and emotion intensity levels. Appearance realism averages were supporting these findings as well although not significantly. Additional analysis showed that emotional scenarios had a significant effect on appeal and emotion intensity perceptions. However, we didn't find any interaction effects between appearance realism and animation realism.

\section{Related work}

A group of previous work investigated the appearance and motion realism of virtual humans but the appearance realism of these characters were not very high. Ferstl et al. \citep{ferstl2021} investigated appearance and gesture motion realism of social agents in general and they compared between a human and an anthropomorphic robot model. They found that the motion condition with highest human-like realism was significantly more likable than all other reduced motion conditions. In this study, the robot model was perceived as significantly more likable than the virtual human, also for the most realistic, human-like gesture motion condition. McDonnell et al. \citep{render_me_real} investigated the effect of different render styles and animation anomalies on the perception of virtual humans. Cartoon renders as well as the most realistic renders were considered significantly more appealing. The render styles categorized in-between these two were perceived as the least appealing. When testing motion anomalies by removing eye-blinking and eye-gaze or turning motion off for half of the face, they found that the anomalies were perceived as unpleasant. On the other hand, a cartoonish render was perceived as the most appealing. These results hint that more realistic virtual humans make people more sensitive towards imperfections. It is to be noted however, that neither of these studies investigated emotionally expressive virtual humans and appearance realism in these studies was not on par with today's standards. Tinwell et al. \citep{TINWELL20131617} removed eyebrow motion in virtual characters and found significant correlation between eeriness ratings and full animation realism and reduced upper face motion. Finally, a number of studies have investigated the perception of emotionally expressive photo-realistic virtual humans, similar to the ones in our study, e.g. Zibrek et al. \citep{10.1145/3349609} and Higgins et al. \citep{higgins_uncanny}. In both studies, appearance realism was manipulated for an expressive virtual human. In the case of Higgins et al., Epic Games' MetaHumans were used, with same appearance realism conditions as in this paper. Both of these studies suggests that the uncanny valley with today's photo-realistic virtual humans has been crossed, as more photorealistic renders were perceived as more appealing in user studies. While previous papers investigated the effects of appearance realism and animation realism separately, none of these studies have investigated these effects at the same time for the photo-realistic characters of today in the context of emotionally expressive scenarios. With this goal in mind, we aim to find the complex relationships between appearance and animation realism including the effect of the emotional tone of the scenarios.

\section{Goal and Hypothesis}

The goal of this study was to investigate the effect of appearance realism (photorealistic, semi-realistic) and animation realism (full, medium, low) on the perceptions towards the virtual human portrayed in the context of different emotional scenarios (neutral, happy, angry, sad). As dependent variables, we measured the level of social presence, affinity, realism and emotion intensity. The term social presence is defined as "being with another" by Biocca et al. \citep{biocca2003toward} and used widely for measurements of perception of virtual humans in VR \citep{photorealism_social_presence} \citep{10.1145/3349609}. Social presense measures are also used for virtual humans in desktop applications as well as for social robots \cite{oh2018systematic}. Towards the goal of quantifying the effect of the uncanny valley, affinity has been introduced \cite{higgins_uncanny} to measure how appealing, eerie and familiar a character appears. Attractiveness is also considered as a metric related to affinity based on the findings of Zibrek et al. \citep{zibrek2020effect} which shows similarities between appeal and attractiveness. Similar to Higgins et al. \cite{higgins_uncanny} we also measure perceived realism in terms of appearance, behaviour and facial movement and added also body movement realism. Finally, we looked into the perception of the emotion intensity \citep{wisessing2020enlighten} to understand the connection between appearance and animation realism and emotion scenarios. While the measures on realism and intensity gives us an indication of how successful our experimental conditions are, social presence and affinity shows the effects of the appearance and animation realism on the perception of users' feeling of being with another being and their level of likeability towards this character. We formulated the following hypothesis:

\begin{itemize}
    \item {\bf H1a and H1b: Photorealistic appearance (a) and full animation realism (b) will lead to significantly higher social presence.}
    \item {\bf H2a and H2b: Photorealistic appearance (a) and full animation realism (b) will lead to significantly higher affinity.}
    \item {\bf H3a and H3b: Photorealistic appearance (a) and full animation realism (b) will lead to significantly higher perceived realism.}
    \item {\bf H4a and H4b: Photorealistic appearance (a) and full animation realism (b) will lead to significantly higher perceived intensity of emotions.}
\end{itemize}

\section{Stimuli creation}

\begin{figure}[!t]
\centering
\includegraphics[scale=.17]{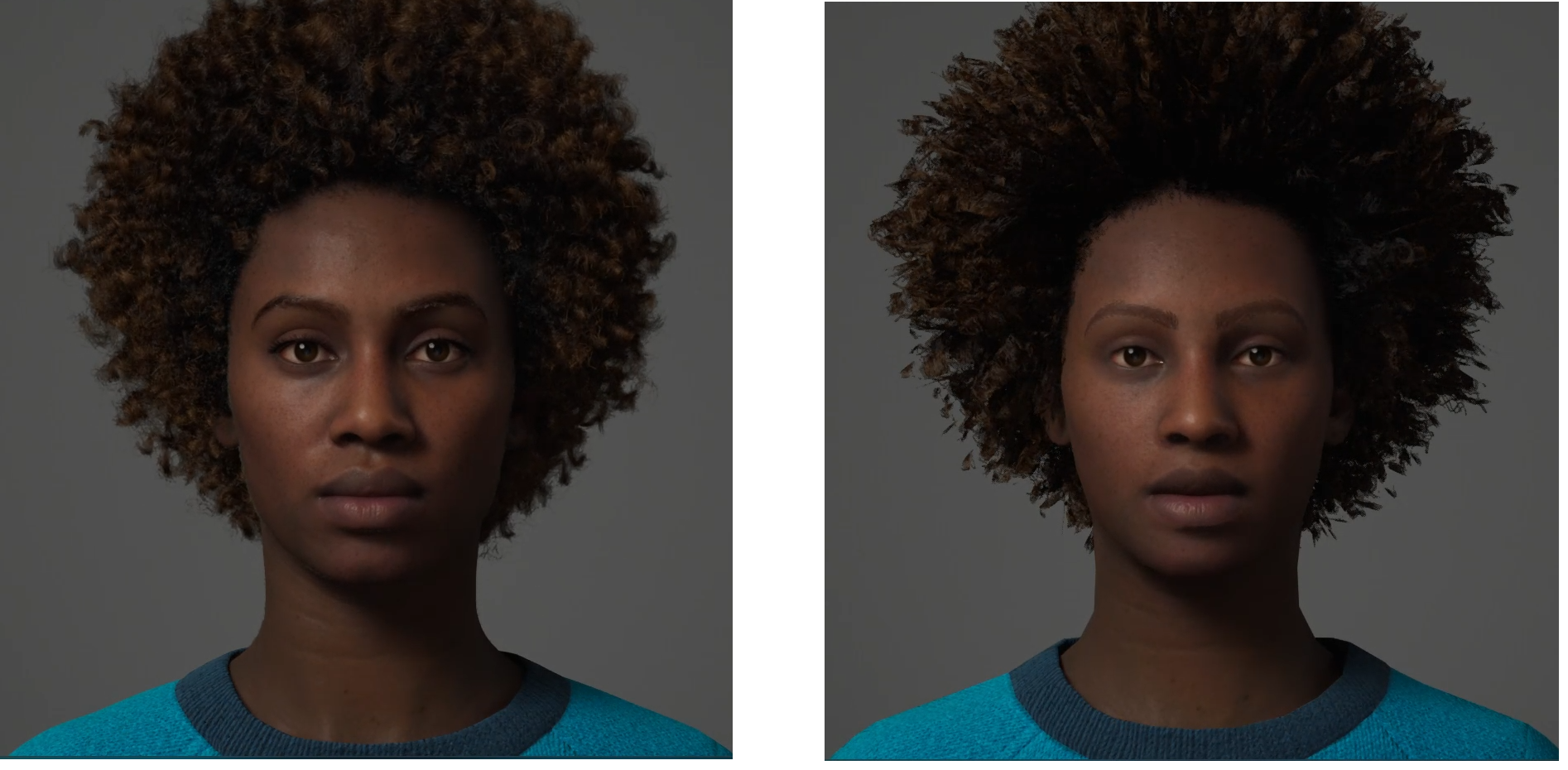}
\caption{Virtual human "Kioko" as used in this study, a MetaHuman released by Epic Games. On the left, the \textit{\textbf{Photorealistic}} version (LOD0). On the right, the \textit{\textbf{Semi-realistic}} version (LOD4).}
\label{fig:appearances}
\end{figure}

To investigate the effect of appearance realism, two appearance realism conditions were created: {\it Photorealistic} and {\it Semi-realistic} using Unreal Engine Metahumans as shown in Figure \ref{fig:appearances}. The {\it Photorealistic} appearance condition was presented with the highest quality level of detail possible which is LOD0. The {\it Semi-realistic} appearance condition was represented at LOD4 to ensure comparability with previous research \citep{higgins_uncanny}. For recording the animations, a commercial video-based facial capture system from Dynamixyz was used and facial animations were retargeted to the MetaHuman. Upper body was also captured using the Vicon motion capture system but only for the purpose of complementing the facial movement with head movement. Facial animation sequences, body animation sequences as well as the voice recordings were rendered in Unreal Engine 5 to produce the video sequences for the on-screen user study (see Figure \ref{fig:teaser}). Four emotional scenarios were designed: {\it Neutral, Happy, Angry} and {\it Sad}. In a previous study, Boaz et al. \citep{emotionSentences} investigated which sentences most correlated with perceived emotions. This was used as a reference to craft 20-second scenario sequences. The control rig in the appearance conditions were exactly the same allowing a fair comparison. As lack of upper face motion has been attributed to higher perceptions of eeriness \citep{TINWELL20131617}, next to the {\it Full} Animation Realism condition, the anomaly conditions for removed eyebrow motion ({\it Medium} Animation Realism) and removed upper face motion ({\it Low} Animation Realism) were created. The video stimuli can be found \href{https://drive.google.com/file/d/1jMVB8tQNFwIaHYkATLrr5L8pY0lvAAvl/view?usp=share_link}{here.}

\section{Experiment Design}
For the user study, participants were recruited on Amazon Mechanical Turk \citep{AMT}. They were guided through the experiment through a survey created on Qualtrics \citep{QualtricsRef}. This survey first displayed a consent form and instructions. Participants were reimbursed 4\$. Each participant watched 12 videos (see Supplementary Material section at the end of the paper). For the two Appearance Realism conditions, there were four emotion scenarios rendered in three animation realism levels. After each video, there was one attention check question, asking a multiple-choice question about elements the virtual human was talking about. Participants' survey submissions were deemed admissible if they passed at least 11 out of 12 attention check questions. This was due to one error being attributed to a genuine mistake amidst 11 correctly answered attention checks (e.g. a mis-click). A total of 94 people took part in the study, however 32 had to be excluded due a number of failed attention checks, leading to 62 participants.  The mean age of participants was 35.58 ({\it SD} = 8.76) with the youngest participant being of age 21 and the eldest being of age 56. On average, their experience with virtual humans or gaming media in general was slightly above average. Gender was balanced with 16 being female and 15 being male in each appearance condition. After being randomly assigned to the {\it Photorealistic} or the {\it Semi-realistic} appearance conditions, participants were then presented with randomly ordered videos displaying different levels of Animation Realism for all Emotion Scenarios. They were asked to rate on a Likert scale from 1 ("Not at all") to 7 ("Extremely") how much they agreed with statements related to our dependent variables. As explained in Section 3, we had 4 categories of dependent variables including social presence, affinity, realism and emotion intensity. An overview of the dependent variables can be found in Table \ref{tab:dep_variables}. 

\begin{table*}[t]
\caption{Dependent variables of this study. Likert scale from 1 ("Not at all") to 7 ("Extremely").}
\centering
\begin{tabular}{p{3cm} p{4cm} p{9cm}}
\hline
Group & Dependent variable & Questionnaire item\\
\hline
\hline \\
Emotion Perception & Intensity & "How intense was the emotion you observed?"\\ \\
\hline \\
 & SP1 & "It feels as if I am in the
presence of another."\\ \\
Social Presence & SP2 & "It feels as if the girl is
watching me and is
aware of my presence."\\ \\
 & SP3 & "The thought that the girl
isn't real crossed my
mind often."\\ \\
 & SP4 & "The girl appears to be
alive."\\ \\
 & SP5 & "The girl is only a
computerized image,
not a real person."\\ \\
 \hline \\
 & Appeal & "I found the girl appealing, likable."\\ \\
 Affinity & Eerie & "I found the girl eerie, creepy."\\ \\
 & Familiar & "I found the girl familiar, I have seen a similar person before."\\ \\
 & Attractive & "I found the girl attractive."\\ \\
 \hline\\
  & Appearance Realism & "I found the girl's appearance realistic."\\ \\
 Realism & Face Movement Realism & "I found the girl's facial movements realistic."\\ \\
  & Body Movement Realism & "I found the girl's body movements realistic."\\ \\
 & Behavior Realism & "I found the girl's behavior realistic."\\ \\
 & Voice Realism & "I found the girl's voice realistic."\\ \\
 & Overall Realism & "I found the girl realistic overall."\\ \\
 \hline
\end{tabular}
\label{tab:dep_variables}
\end{table*}

\section{Results}

A mixed three-way ANOVA design was used with between-subject factor Appearance Realism and within-subject factors Animation Realism and Emotion Scenario. In case of significance, post hoc analysis were done with Bonferroni pairwise comparisons. The homogeneity of variance assumption was tested using Levene's test, while the assumption of sphericity was checked with Mauchly's test. In cases where the homogeneity of variance assumption was violated, a non-parametric equivalent of a mixed three-way ANOVA was used \cite{JSSv050i12}. In case of statistically significant effects, non-parametric post-hoc pairwise comparisons followed \cite{JSSv064i09}.

%An overview of results can be found in Table \ref{tab:significant} and \ref{tab:significant2}. In the following subsections, they are described in detail.

\subsection{Social Presence (H1a and H1b)}
%Keep it consistent:
%test for assumptions
%Three-way ANOVA
%Two-way ANOVA
%post hoc
%extra tests at the end
\begin{figure}[!t]
\centering
\includegraphics[scale=.135]{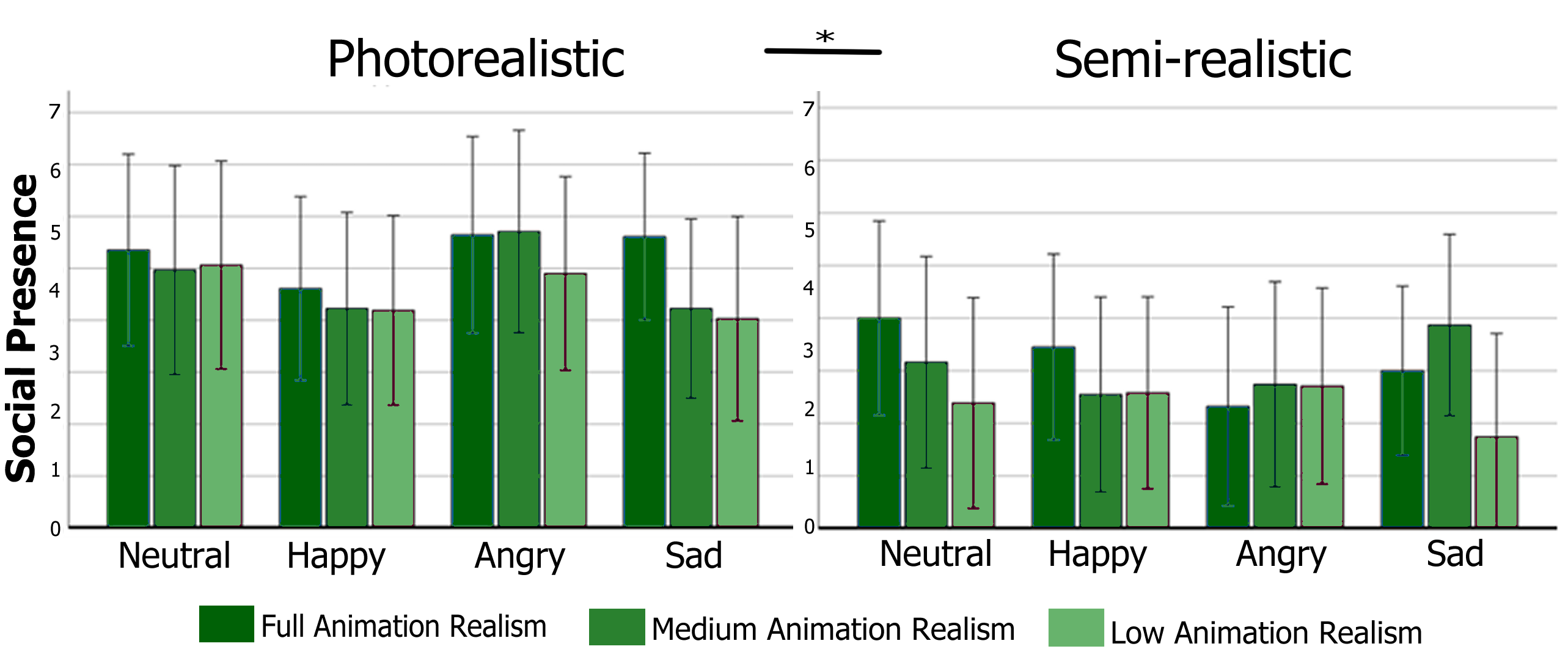}
\caption{Mean values for the variable Social Presence. }
\label{fig:SP_all}
\end{figure}
For the independent variable Appearance Realism, a three-way ANOVA revealed a significant main effect ($F(1,60) = 4.057, p = 0.048$). As seen in Figure \ref{fig:SP_all}, there was a noticeably different Social Presence rating for the appearance condition {\it Photorealistic} (estimated $\overline x$ = 4.879, { \it SE} = 0.703) as compared to the {\it Semi-realistic} condition (estimated $\overline x$ = 2.876, { \it SE} = 0.703), indicating that social presence was indeed higher for the photorealistic appearance condition. 

Figure \ref{fig:SP_all} shows noticeable differences between the three levels, {\it Low} (estimated $\overline x$ = 3.452, {\it SE} = 0.559), {\it Medium} (estimated $\overline x$ = 3.931, {\it SE} = 0.547) and {\it Full} Animation Realism (estimated $\overline x$ = 4.25, {\it SE} = 0.499). However, no significant main effect was found for the variable Animation realism. Similarly, no statistical significance was found for Emotion Scenario. Here, the means showed only slight differences for the scenarios Neutral ($\overline x$ = 4.156, {\it SE} = 0.629), Happy ($\overline x$ = 3.602, {\it SE} = 0.559), Angry ($\overline x$ = 4.005, 22 {\it SE} = 0.591) and Sad ($\overline x$ = 3.747, {\it SE} = 0.504). Finally, no interaction effects were found. 

\subsection{Affinity (H2a and H2b)}

\paragraph{Appeal}
%{\bf Appeal}
%say that even though we investigted hypotheses mainly, we also looked at X
 Neither Animation Realism, nor Appearance Realism had any main effects that proved significant for the measure Appeal. Furthermore, no statistically significant interaction effects were observed. Overall, the Appeal ratings were above or around average. {\it Photorealistic} Appearance (estimated $\overline x$ = 4.223, {\it SE} = 0.238) scored slightly higher than {\it Semi-realistic} Appearance (estimated $\overline x$ = 4.14, {\it SE} = 0.238) on the Appeal scale. Similarly, higher Animation Realism consistently resulted in higher Appeal - with {\it Full} Animation Realism (estimated $\overline x$ = 4.286, {\it SE} = 0.179) being more appealing than {\it Medium} Animation Realism (estimated $\overline x$ = 4.19, {\it SE} = 0.16) and {\it Medium} Animation Realism being more appealing than {\it Low} Animation Realism (estimated $\overline x$ = 4.069, {\it SE} = 0.19). Although, there are no significant effects as related to the hypotheses, a significant effect of Emotion Scenario ($F(2.586, 155.16) = 10.976, p < 0.001$) was observed. Subsequent post hoc pairwise comparisons highlighted that the Emotion Scenarios {\it Angry} and \textit{Sad} scored significantly lower than the Emotion Scenarios {\it Neutral} and {\it Happy} (see Table \ref{tab:significant}).
 
\begin{figure}[!t]
\centering
\includegraphics[scale=.135]{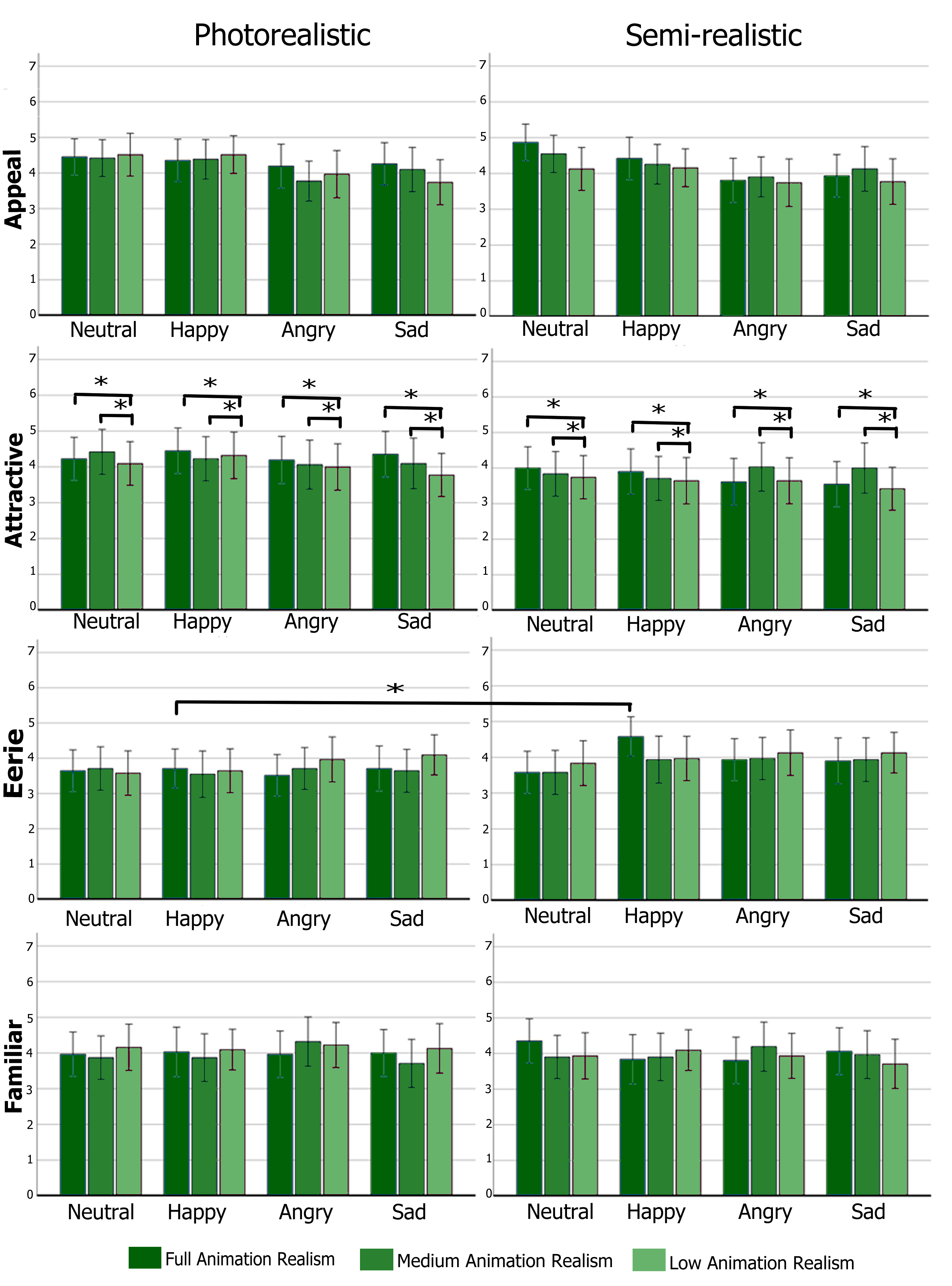}
\caption{Mean values for affinity.}
\label{fig:all_affinity}
\end{figure}

\paragraph{Eerieness}
 No statistically significant main or interaction effects were found for appearance and animation realims. However, in Figure \ref{fig:all_affinity}, a tendency can be observed for appearance realism. {\it Semi-realistic} appearance (estimated $\overline x$ = 3.957, SE = 0.218) reveals overall higher eerieness scores than {\it Photorealistic} appearance (estimated $\overline x$ = 3.707, SE = 0.218). In addition, a tendency for animation realism can be observed as well. Overall, {\it Low} animation realism leads to higher eerieness scores (estimated $\overline x$ = 3.919, {\it SE} = 0.177) than {\it Full} animation realism (estimated $\overline x$ = 3.823, {\it SE} = 0.168). There was a noticeable peak in the \textit{Happy} emotion in the semi-realistic appearance condition acting against the trend however. 

 \paragraph{Familiarity}
No statistically significant main or interaction effects or tendencies that supports the hypothesis could be observed for the measure familiarity.

\paragraph{Attractiveness}
 There was a statistically significant main effect for Animation Realism ($F(1.93, 115.8) = 3.527, p = 0.03$). Subsequent post hoc pairwise comparisons revealed that the {\it Low} Animation Realism condition scored significantly lower on the attractiveness scale than the {\it Medium} ($p = 0.025$) and {\it Full} ($p = 0.036$) animation realism conditions. No statistically significant main effect of Appearance Realism could be found. However, a general tendency can still be observed in Figure \ref{fig:all_affinity}. Attractiveness scores for {\it Photorealistic} appearance reveal an overall higher score (estimated $\overline x$ = 4.185, {\it SE} = 0.277) than {\it Semi-realistic} appearance (estimated $\overline x$ = 3.758, {\it SE} = 0.277).

\subsection{Realism Measures (H3a and H3b)}

\begin{figure}[!t]
\centering
\includegraphics[scale=.135]{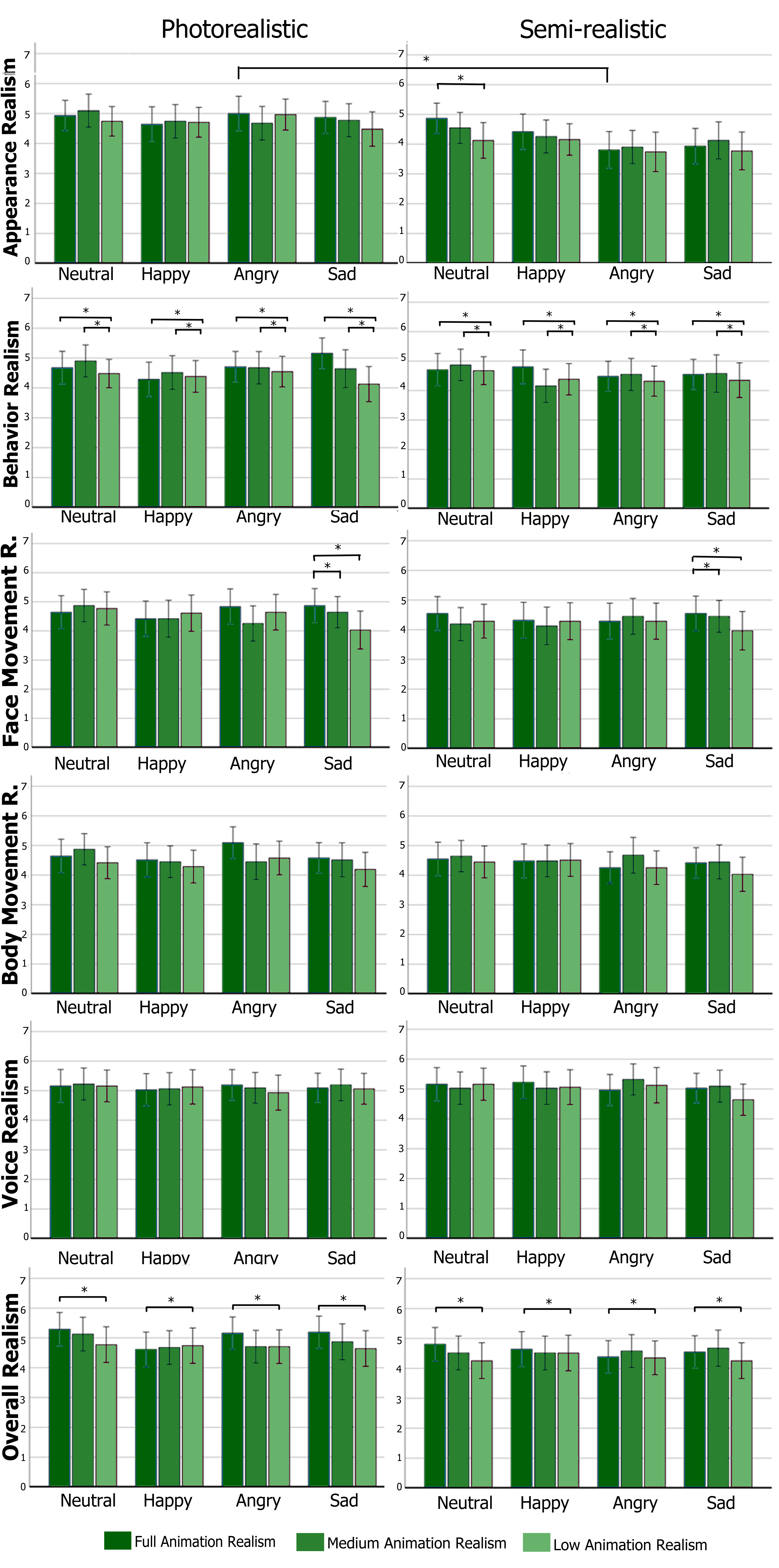}
\caption{Mean values for realism.}
\label{fig:realism_all}
\end{figure}

\paragraph{Appearance Realism }
There were no significant main or interaction effects of appearance and animation realism conditions on perceived appearance realism. However, Figure \ref{fig:realism_all} shows that the estimated averages for the {\it Photorealistic} appearance condition ($\overline x$ = 4.8, {\it SE} = 0.203) is higher than the {\it Semi-realistic} appearance condition ($\overline x$ = 4.478, {\it SE} = 0.203). The results give indication that the {\it Photorealistic} appearance realism condition was perceived better with an estimated mean clearly above average, however with room for improvement.

\paragraph{Face Movement Realism }
No statistically significant main effects were found for animation or appearance realism. Yet, a statistically significant interaction effect of animation realism and emotion scenario was found ($F(5.597, 335.841) = 2.234, p = 0.044$). Post hoc pairwise comparisons with Bonferroni 95\% confidence intervals highlighted that the face movement realism scores was lower for the condition {\it Low} as compared to the scores for the {\it Medium} ($p = 0.024$) and {\it Full} ($p = 0.007$) Animation Realism conditions. The results hint that the {\it Full} Animation Realism condition had a higher perceived face movement realism. This gave some hints that the animations generated were well-received. With an estimated $\overline x = 4.56$ ({\it SE} = 0.162), the animation quality was overall deemed above average. However, there is still room for more realistic facial animations.

\paragraph{Behavior realism }
%For the measure Behavior Realism, Levene's test highlighted one sample group that violated the homogeneity of variances assumption ($p = 0.005$). Therefore, a non-parametric counterpart to the three-way mixed ANOVA was employed. 
A significant main effect of animation realism was found ($F(1.977,118.62) = 3.23, p = 0.04$). Post hoc %non-parametric
pairwise comparisons showed that the {\it Low} Animation Realism condition led to significantly lower ratings for behavior realism than the conditions {\it Medium} ($p = 0.04$) and {\it Full} ($p = 0.015$).

\paragraph{Overall realism }
For the measure overall realism, there was a statistically significant main effect of Animation Ralism ($F(1.889, 113.34) = 4.417, p = 0.014$). Subsequent post hoc pairwise comparisons showed that the {\it Low} Animation Realism condition led to significantly lower overall realism ratings than the {\it Full} Animation Realism condition ($p = 0.004$).

\paragraph{Voice Realism and Body Movement Realism}
In line with expectations, no statistically significant main or interaction effects could be observed for the measures voice realism and body movement realism as we were only changing the facial animations and appearance but not the body movement or the voice.

\subsection{Intensity (H4a and H4b)}
\begin{figure}[!t]
\centering
\includegraphics[scale=.135]{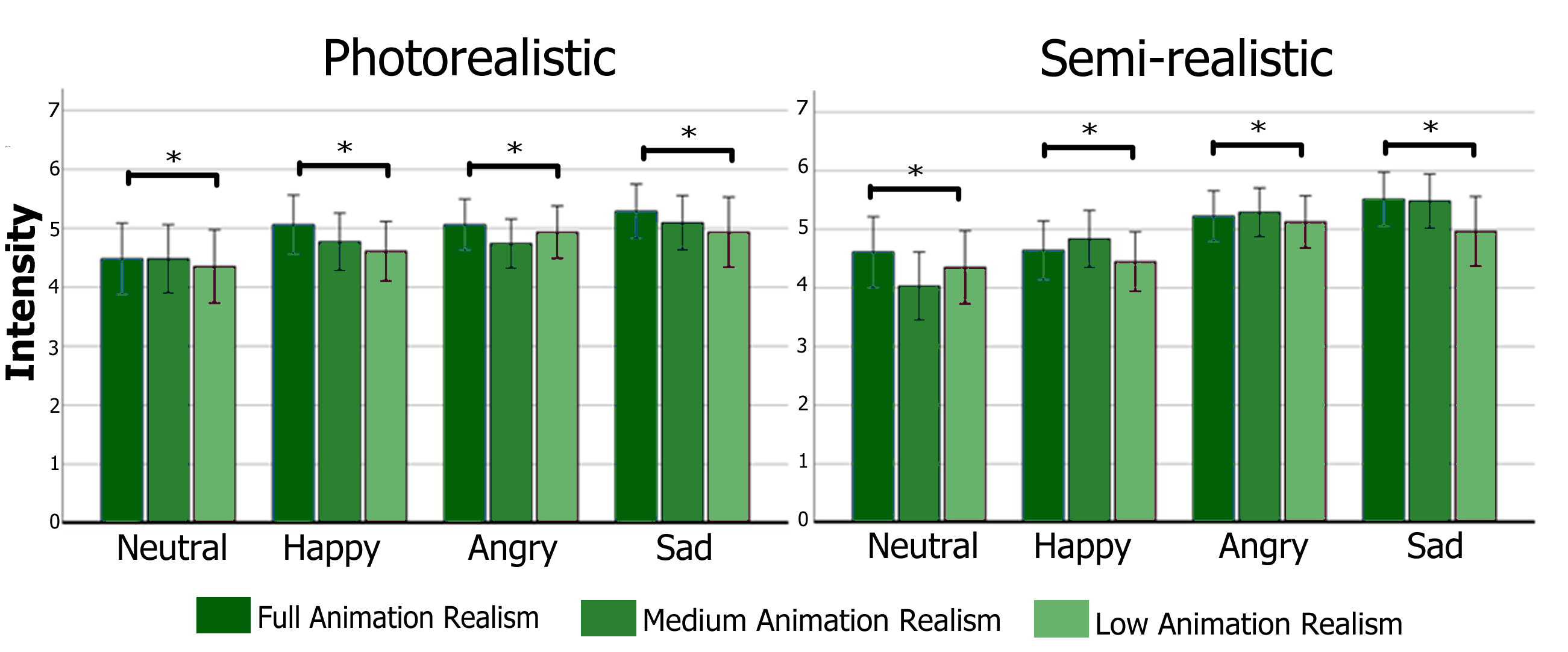}
\caption{Mean values for intensity.}
\label{fig:Intensity}
\end{figure}

No statistically significant main or interaction effects for Appearance Realism was found. However, as shown in Figure \ref{fig:Intensity}, estimated means for the {\it Full} (estimated $\overline x = 4.988$, {\it SE} = 0.127), {\it Medium} (estimated $\overline x = 4.843$, {\it SE} = 0.133) and {\it Low} (estimated $\overline x = 4.718$, {\it SE} = 0.144) Animation Realism conditions shows that higher animation realism is perceived with higher emotion intensity. There was also a statistically significant main effect of emotion scenario ($F(2.699,161.97) = 11.756, p < 0.001$). Post hoc pairwise comparisons showed that participants perceived the {\it Sad} Emotion Scenario to be significantly more intense than the {\it Neutral} ($p < 0.001$) and {\it Happy} ($p < 0.001$) Emotion Scenarios. Similarly, the {\it Angry} emotion was perceived to be significantly more intense than the emotion scenarios {\it Neutral} ($p < 0.001$) and {\it Happy} ($p = 0.013$). That also shows indications for the quality of the design of the emotion scenarios.

% keep it consistent for separate testsfor emotion conditions
\begin{table*}[!t]
\caption{\label{tab3}Statistical Significance}
\centering
\begin{tabular}{p{6.4cm} p{4cm} p{6.5cm}}
\hline
Effect & ANOVA Result & Post hoc\\
\hline
Social Presence & &\\
\hline
SOCIAL PRESENCE: Appearance Realism &$F(1, 60) = 4.057, p = 0.048$& Photorealistic appearance leads to higher social presence than semi-realistic appearance ($p = 0.048$) \\
\hline
Affinity & &\\
\hline
APPEAL: Emotion Scenario & $F(2.586, 155.16) = 10.976$, $p < 0.001$& Emotion Scenario{ \it Angry} significantly less appealing than {\it Neutral} ($p < 0.001$) and {\it Happy} ($p < 0.001$). Emotion Scenario {\it Sad} also significantly less appealing {\it Neutral} ($p = 0.042$) and {\it Happy} ($p = 0.027$) scenarios.\\
\\
ATTRACTIVENESS: Animation Realism & $F(1.93, 115.8) = 3.527$, $p = 0.03$ & \textit{Low} Animation Realism significantly less attractive than \textit{Medium} ($p = 0.025$) and \textit{Full} Animation Realism ($p = 0.036$).\\
\hline
Realism & &\\
\hline

FACE MOVEMENT REALISM: Animation Realism x Emotion Scenario & $(F(5.597, 335.841) = 2.234, p = 0.044)$ & \textit{Low} Animation Realism perceived to be significantly less realistic than \textit{Medium}
($p = 0.024$) and \textit{Full} ($p = 0.007$) Animation Realism for the Sad Emotion Scenario\\
\\
BEHAVIOR REALISM: Animation Realism & $F(1.977, 118.62) = 3.23$, $p = 0.04$ &\textit{Low} Animation Realism led to significantly lower perceived behavior realism than \textit{Medium} ($p = 0.04$) and \textit{Full}
($p = 0.015$)\\
\\
OVERALL REALISM: Animation Realism & $F(1.889, 113.34) = 4.417$, $p = 0.014$ &\textit{Low} Animation Realism led to significantly lower perceived overall realism than \textit{Full} ($p = 0.004$)\\
\hline
Intensity & &\\
\hline
INTENSITY: Emotion Scenario &$F(2.699, 161.97) = 11.756$, $p < 0.001$& Emotion Scenario \textit{Sad} perceived as significantly more intense \textit{Neutral} ($p <0.001$) and \textit{Happy} ($p < 0.001$) scenarios. Emotion Scenario \textit{Angry} perceived as significantly more intense than scenarios \textit{Neutral} ($p < 0.001$) and \textit{Happy} ($p = 0.013$).\\
\\
INTENSITY: Animation Realism & $F(1.920,115.219) = 3.225$, $p = 0.045$ & \textit{Full} Animation Realism significantly more intense than \textit{Low} Animation Realism ($p = 0.025$).\\
\hline
\end{tabular}
\label{tab:significant}
\end{table*}

\section{Discussion}
Results highlighted that appearance realism had a significant effect on social presence. That was inline with the findings of Zibrek et al. \citep{photorealism_social_presence} \citep{zibrek2017don} but it was shown with state-of-the-art photorealistic virtual humans in desktop applications in our case. Past studies investigating social presence were using characters more comparable in appearance realism to the semi-realistic character in this study. The results shows the increased level of social presence with higher appearance realism building upon the previous results. Furthermore, while there was no significant main effect of animation realism on social presence, there were signs that higher animation realism led to higher social presence.
%if study found before, any differences? no difference? okay great - if differences, explain why
%focus on significances, not on what wasn't significant
%end with something good, don't just report for the sake of reporting, make a point, arguef or your paper
With regards to appeal, % For appeal, on the one hand, it supports a recent study that suggests that we are at a stage where virtual humans are photorealistic enough for animation anomalies to not matter anymore, on the other hand, even for the semi-realistic appearance, no significant influence on appeal could be observed - only emotion scenario significantly affected appeal. However, 
 while no statistical significance for animation realism could be found, the results show a clear tendency where higher animation realism led to higher appeal of the virtual character. Additionally, data revealed that the photorealistic condition was perceived as more appealing than the semi-realistic condition, although not significantly. %This suggests that for virtual humans to appear more appealing in one's applications, for instance for ads involving virtual humans, one should aim towards scenarios with neutral or happy settings. Additionally, an overall trend can be observed as the semi-realistic character led to a clear higher average as compared to the photorealistic character. 
 %In conclusion, the state-of-the-art photorealism possible today lends itself well to such tasks as they are seen as more appealing.\\ 
We found out that {\it Low} Animation Realism and {\it Semi-realistic} Appearance conditions have higher eeriness scores which was inline with our initial hypothesis, however the effect was not significant. Previous research by Tinwell et al.\citep{TINWELL20131617} showed similar results however they were using additional external stimuli such as sound and environment. We conclude that whole face animation is preferable in comparison to anomaly conditions. While the removal of eyebrow motion didn't lead to significantly different eerie ratings, there was a significant change in attractiveness perceptions. Furthermore, changing animation realism led to a significant main effect for perceived differences in behavior and overall realism. Therefore, more realistic motions are perceived as more attractive which supports findings by Zibrek et al. \citep{zibrek2020effect} \citep{zibrek2022proximity} but different from previous work it has been shown with highly realistic characters. 
 
Finally, the emotion intensity was perceived differently based on emotion scenario. The angry and sad emotion scenarios were perceived to be a lot more intense than other scenarios. That highlights that appearance and animation realism can become even more important in different emotionally expressive situations. Another observation that is poignant is that the emotions were perceived as significantly less intense for the {\it Low} Animation Realism condition which reveals the importance of Animation Realism. We were expecting upper face motion to be very important in portraying and displaying emotion. Complete removal of upper face motion did indeed led to significantly lower intensity ratings. However, the lack of eyebrow motion only didn't lead to significantly different perceptions of emotion intensity. While the averages for the different animation realism conditions do show a difference between all three levels of animation realism, the influence of eyebrow motion didn't lead to differences that were big enough to warrant statistical significance. This might be due to the short length of the video stimuli as the scenes were limited to around 20 seconds. Additionally, the eyebrow motion of humans is quite subtle which might explain this effect. Additionally, we only used the intensity of the emotions in our measures and didn't take the emotion label into account. More complex self-report measures can be used such as \cite{scherer2013grid}.

\section{Limitations and future work}

Although the results reveal some evidence on the effect of appearance and animation realism on the perception of emotionally expressive virtual characters, there are several open issues and limitations. The experiment was conducted with only one female virtual human model with a particular identity. It can be interesting to investigate if and how people would react differently to other virtual humans with regards to ethnic background and gender. The lip sync animation was not perfect and has room for improvement. Lip sync also leads to the question in how far the effect of voice is prevalent in the perception of virtual humans. Although we asked the users' about how they perceive the effect of voice realism and found no significant effect on the study, as our perception is naturally cross-modal \citep{stein1993merging}, voice has a definite effect on perception as also shown in previous research \citep{HIGGINS2022116}\citep{brambilla2021faces}. A follow-up study could investigate in how far simultaneous manipulation of voice realism has an influence on the perception of the virtual human. Moreover, in this study appearance realism changes were achieved by changing the level of detail of photorealism level in Metahumans. There are further comparisons to be made, for example comparing different type of photorealistic characters with varying level of photorealism not only represented in LOD differences. So the definition of semi-realistic appearance is much broader than a change in LOD. Similarly, animation realism in this study was manipulated in a specific way - manipulations with more granularity and more variation might lead to new or different insights. For example instead of activating/deactivating certain parts of facial animation, the overall intensity and complexity of the facial animation can be changed (number of blendshapes, additional effects such as wrinkles etc). In addition, a large participant size could be aimed for which can lead to improvements in the significance of the results. Furthermore, this study was limited to an on-screen experiment and more experiments are needed in immersive environments with VR headsets. In this study, the background of the virtual human was limited to a blank screen. It can also be an interesting research question to investigate the effect an environment has on the perception of an emotionally expressive virtual human. Additionally, this research didn't allow for a responsive virtual human which could engage and react to participants. It was limited to an on-screen video presentation. This gives rise to the question how the realism perception of virtual humans is influenced in an interactive setting. 

To conclude, our study sheds light into how appearance and animation realism effects the perception of highly realistic virtual humans in emotionally expressive scenarios. We hope that these findings provide input to follow up studies in this area.

\balance
%\section*{Supplementary Material}

%Supplementary material is added in the form of video renders with the virtual human as presented to participants. It can be found here: %\url{https://drive.google.com/file/d/1jMVB8tQNFwIaHYkATLrr5L8pY0lvAAvl/view?usp=share_link}
%\begin{acks}
%To Robert, for the bagels and explaining CMYK and color spaces.
%\end{acks}

\bibliographystyle{ACM-Reference-Format}
\bibliography{refs}

\end{document}